\documentclass[%
 reprint,
showpacs,
 amsmath,amssymb,
 aps,
]{revtex4-1}

\usepackage{graphicx}
\usepackage{dcolumn}
\usepackage{bm}

\begin{document}

\title{Ballistic Metamaterials}

\author{ Kun Li$^{1*}$, Evan Simmons$^{2*}$, A. F. Briggs$^1$, S. R. Bank$^1$, Daniel Wasserman$^1$, Viktor A. Podolskiy$^2$ and Evgenii E. Narimanov$^3$   }


\affiliation{${}^1$ Electrical and Computer Engineering Department, University of Texas, Austin TX 78712, USA} 
\affiliation{${}^2$ Department of Physics and Applied Physics, University of Massachusetts Lowell, Lowell, MA 01854, USA} 
\affiliation{${}^3$ School of Electrical and Computer Engineering,  and Birck Nanotechnology Center, Purdue University, West Lafayette, IN 47907, USA}

\thanks{These authors  contributed equally to this work.}

\maketitle

{\bf   The interaction of free electrons with electromagnetic excitation is the fundamental mechanism responsible for ultra-strong confinement of light \cite{Cai2009} that, in turn, enables biosensing, near-field microscopy,\cite{NFM} optical cloaking,\cite{cloaking-proposal-Pendry,cloaking-proposal-Leonhardt,cloaking-expt,cloaking-optical} sub-wavelength focusing,\cite{hyper-focusing-Viktor,hyper-focusing-expt,Natasha} and super-resolution imaging.\cite{Pendry2000,hyperlensEN,hyperlensNE,hyperlens-Zhang,hyperlens-3D} These unique phenomena and functionalities critically rely on the negative permittivity of optical elements \cite{negative-refraction-expt} resulting from the free electrons. As result, progress in nanophotonics and nano-optics is often related to the development of new negative permittivity (plasmonic) media at the optical frequency of interest.\cite{Cai2009}
Here we show that the essential mobility of free charge carriers in such conducting media dramatically alters the well-known optical response of free electron gases. 
We demonstrate that a ballistic resonance associated with the interplay of the time-periodic motion of the free electrons in the confines of a sub-wavelength scale nanostructure and the time periodic electromagnetic field leads to a dramatic enhancement of the electric polarization of the medium --  to the point where a plasmonic response can be achieved in a composite material using only positive bulk permittivity components.  
This ballistic resonance opens the fields of plasmonics, nanophotonics, and metamaterials to many new constituent materials that until now were considered unsuitable for such applications, and extends the operational frequency range of existing materials to substantially shorter wavelengths.  As a proof of concept, we experimentally demonstrate that ballistic resonance in all-semiconductor metamaterials results in strongly anisotropic (hyperbolic) response well above the plasma frequencies of the metamaterial components. }

These ballistic semiconductor metamaterials offer a practical path to leverage the maturity of the semiconductor industry to revolutionize photonics and optoelectronics. Furthermore, since the  ballistic resonance, theoretically predicted and experimentally demonstrated in the present work using an all-semiconductor metal/dielectric materials system, is fundamentally different from the effect of inter-subband transitions in quantum wells -- these two effects can be combined within the same material platform to further enhance the electromagnetic response as well as incorporate active components, and thus offer even more control over the optical properties of the resulting composite. Finally, the effect demonstrated here in planar layered metamaterials can be extended to other confined geometries, including wire- and particle-based nanostructures and composites.

\section*{Ballistic Resonance}

The optical response of a wide class of materials that includes metals, semi-metals, transparent conducting oxides, and doped semiconductors, is dominated by the dynamics of their free electrons. In a composite (meta)material that include both conducting and dielectric constituents, the resulting free carrier contribution to the electromagnetic response of the system is defined by the relation of the electron  de Broglie wavelength $\lambdabar_F$ at the Fermi energy and the characteristic dimensions of the metallic component of the metamaterial unit cell $d$. 

\begin{figure*}[htbp] 
   \centering
    \includegraphics[width=7. in]{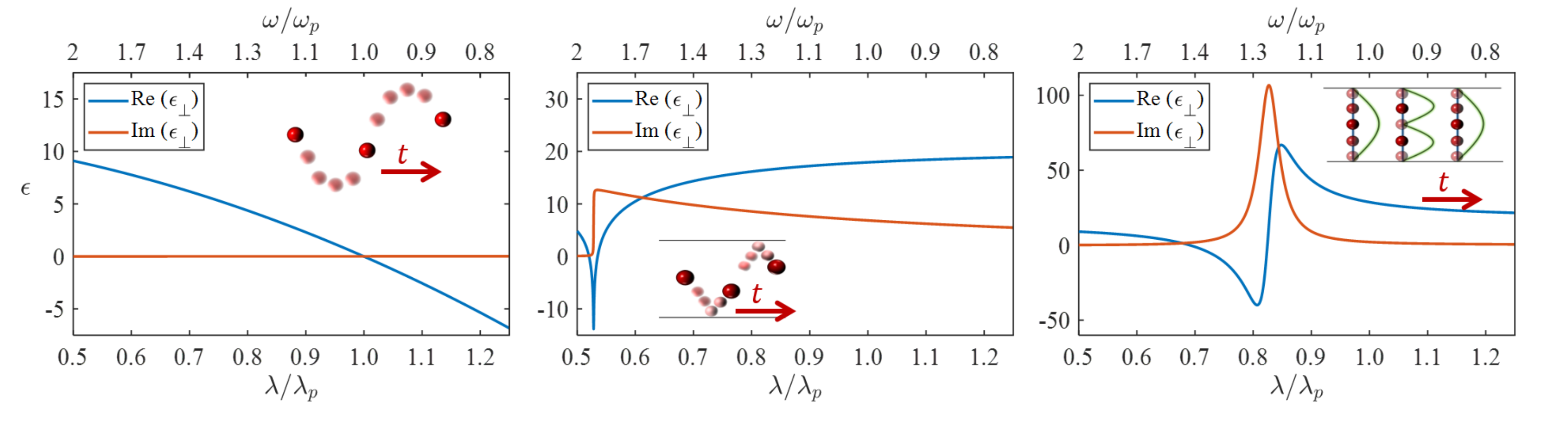}
     \caption{Free-electron-dominated optical response of plasmonic material operating in (a) unconstrained free-electron Drude regime, (b) ballistic  regime, and (c) quantum regime; optical response of thin layers is anisotropic with confinement affecting only motion of electrons perpendicular to layers; insets illustrate electron motion as a function of time. Note the drastic difference in the spectral behavior of permittivity in the three regimes. }
     \label{fig:1}
\end{figure*}

When $\lambdabar_F$ is comparable to $d$, the free electron motion in the conducting elements of the metamaterial unit cell is strongly quantized,\cite{ref:quantum}  leading to the characteristic Lorenzian profile (see Fig. \ref{fig:1}(c)) centered at the frequency of the corresponding quantum transition between different energy levels. In particular, a  planar  metamaterial operating in the regime $\lambdabar_F \sim d $ is essentially a multiple quantum well system, with the resonances in the effective permittivity  corresponding to the electronic inter-subband transitions.\cite{QCL,sinclair,genevet,zayatsNb} 

In the opposite limit $\lambdabar \ll d $, the quantum interference effects can be neglected, and the free electron response is generally treated within the framework of the standard Drude model,\cite{Cai2009} leading to the effective local permittivity 
\begin{eqnarray}
\epsilon_M(\omega)=\epsilon_\infty\left(1-\frac{\omega_p^2}{\omega(\omega+i/\tau)}\right), \label{eq:Drude}
\end{eqnarray}
where $\omega_p$ is the corresponding plasma frequency and $\tau$ is the free electron  scattering time, while  $\epsilon_\infty$ represents  the contribution of the bound electrons. The characteristic frequency dependence of $\epsilon_M\left(\omega\right)$ is shown in Fig. \ref{fig:1}(a).

However, in the semiclassical regime $\lambdabar \ll d $, the actual dynamics of the free electrons can be either 
diffusive \cite{Ziman,ZimanP}  -- when each electron undergoes multiple collisions before reaching the metal-dielectric interface, or 
ballistic \cite{ZimanP,Datta} when the electron mean free path $\ell \gg d$. In the diffusive regime,\cite{ZimanP}  the standard description in terms of the local dielectric permittivity is generally adequate,\cite{hyperlensEN} and the finite unit cell size corrections can be incorporated via the spatial dispersion formalism.\cite{LL,Viktor}
The ballistic regime, on the other hand, shows qualitatively different behavior, with the resonant response due to the interplay of the electromagnetic field period $2\pi/\omega$ and the ballistic round-trip time $2 d/v_F$, where $v_F$ is the Fermi electron velocity. As we show in the present work, when the round-trip time of the electron motion is equal to the period of the electric field, the resulting electromagnetic response is enhanced, leading to the negative effective permittivity in the direction normal to the metal-dielectric interface, even above the plasma frequency when the corresponding bulk permittivity is positive -- see Fig. \ref{fig:1}(b). { Note that this {\it ballistic resonance} does not involve the quantum interference of the electron de Broglie waves, and is therefore qualitatively different from the inter-subband oscillations in quantum wells. }

 \begin{figure*}[htbp] 
   \centering
    \includegraphics[width=7. in]{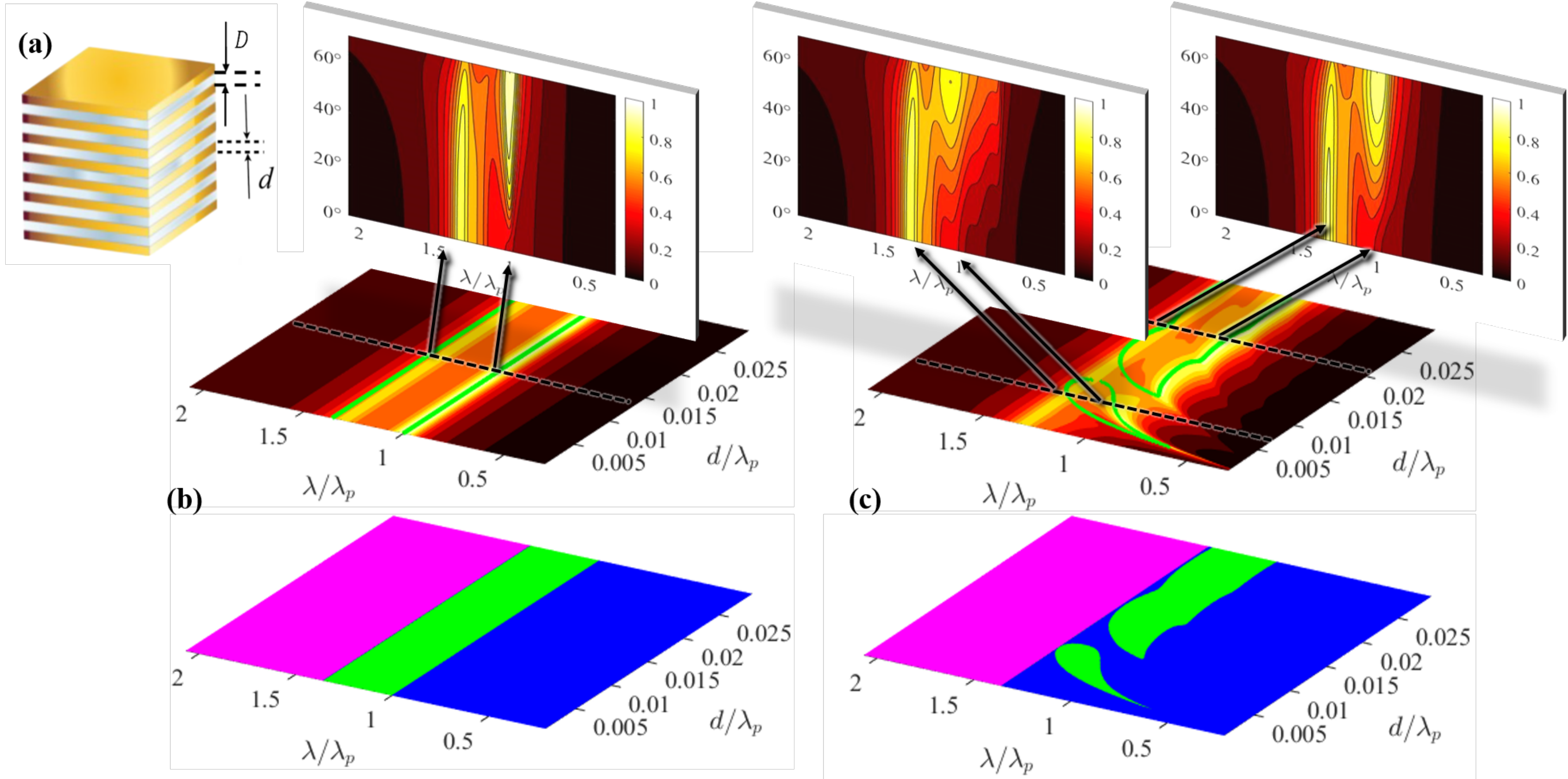}
     \caption{
    (a) the schematic of the ballistic metamaterial and (b,c) ``phase diagrams'' for the  planar layered meta-materials, with equal volume fractions of plasmonic and dielectric ($\epsilon_d=10.23$) components calculated using the effective medium theory (b) and the semiclassical approximation (c). The hyperbolic phase with ${\rm Re}\left[\epsilon_\tau\right] < 0$, ${\rm Re}\left[\epsilon_n\right] > 0$  is represented by magenta color,  the hyperbolic phase  with ${\rm Re}\left[\epsilon_\tau\right] > 0$, ${\rm Re}\left[\epsilon_n\right] < 0$ is shown in green, and elliptic (dielectric) phase -- with blue color. Insets in (b,c) illustrate the expected behavior of absorption profiles as functions of layer size and incident angle; note that the onset of both hyperbolic phases corresponds to the absorption resonances.}
     \label{fig:2}
\end{figure*}

To uncover the key features of (meta-)material response in the ballistic regime, we will focus on the planar geometry of Fig. \ref{fig:2}(a).  In this case the confinement  primarily affects the motion of electrons in the direction that is perpendicular to the interface, and the effective permittivity of the conducting layer becomes anisotropic (see Methods for details). For a high-quality interface when the free electron is reflected from the boundary with essentially the same in-plane component of its momentum, 
the electromagnetic response along the layer can be represented by the effective permittivity that is still described by the conventional Drude model, 
\begin{eqnarray}
\epsilon^{\rm eff}_\tau & = & \epsilon_M\left(\omega\right).
\label{eq:et}
\end{eqnarray}
However, as the reflection from the interface inevitably changes the normal component of the electron velocity,  in the direction normal to the interface the free carrier electromagnetic response  is strongly modified.
When the layer is substantially thick ($d\gg\lambdabar_F$), the effective permittivity in the direction perpendicular to the layer can be calculated using the semiclassical approach, based on the self-consistent solution of the Boltzmann kinetic equation \cite{Ziman} for the charge carrier distribution function in the layer, in the electromagnetic field defined by Maxwell's equations with charge and current densities that, in turn, depend on the electron distribution function \cite{ZimanP} (see {\it Appendix C}), resulting in 
\begin{eqnarray}
\epsilon^{\rm eff}_n & = & \epsilon_M\left(\omega\right) - \epsilon_\infty \cdot \frac{ \omega_p^2}{\omega \left(\omega + i / \tau\right)}
 \cdot F_z\left(\frac{d\left(  \omega + i / \tau \right)}{2 v_F } \right), \ \ \ \  \label{eq:ensc}
\end{eqnarray}
where $v_f$ is the Fermi velocity of electrons and the function $F_z$  is defined as
\begin{eqnarray}
F_z\left(x\right) & = & \frac{3}{x} \int_1^\infty dt \cdot \frac{\tan\left(t x\right)}{t^5}, \label{eq:Fz}
\end{eqnarray}
giving a logarithmic singularity in the low-loss limit when $d = \left(2 n + 1\right) \pi v_F/\omega$ for an integer $n$, corresponding to the ballistic resonance. Fig. \ref{fig:1}(b) shows the resulting frequency dependence of the  effective permittivity normal to the layer $\epsilon_n^{\rm eff}\left(\omega\right)$. Note the difference of the frequency profile of the ballistic resonance from the usual Lorentzian shape: here it's the real part of the permittivity which peaks at the resonance, while the imaginary component  which accounts for loss, rapidly drops to near-zero in a step-like fashion (see {\it Appendix F}).

The physical origin of the non-zero imaginary part of the permittivity in the low-loss limit $1/\tau \to 0$ is the Landau damping.\cite{LL}  For simplicity, consider the frequencies near the primary ($n = 0$) ballistic resonance ($\omega_0 = v_F / \pi d$).  There, an electron with the velocity $v_z = d \omega / \pi$  exhibits periodic motion in the $z$ direction (normal to the interface), with temporal period $2 d/v_z = 2 \pi / \omega$ equal to that of the electromagnetic field. As a result, 
such a free electron  moves ``in phase'' with the electromagnetic field, and can therefore continuously  absorb the field's energy.
For a degenerate Fermi gas, $v_z$ has to be below the Fermi velocity $v_F$, so that such in-phase motion is only possible when $\omega < v_F / \pi d$, i.e. only at the frequencies below but not above the ballistic resonance -- which explains the asymmetric absorption profile.

When the layer thickness $d\sim\lambdabar_F$, the semiclassical approach becomes inadequate and the optical response should be based on full quantum-mechanical calculations. In the limit of ultra-thin layers, the optical response is dominated by discrete quantum well transitions, recovering the expected Lorentzian line shapes (Fig.1c). Importantly, the results of full quantum-mechanical calculations agree with predictions of semiclassical theory for relatively thick layers. Likewise, semiclassical theory predicts $\epsilon^{\rm eff}_n\to \epsilon_M$ as $d/\lambdabar_F\to \infty$. 

The approach derived in this section is also directly applicable to curved layers as long as the radius of curvature is much larger that the layer thickness $d$, and it can be re-formulated to describe optical properties of wires and nanoparticles of various shapes. In the latter case, we expect the results to be shape-dependent.

\section*{Ballistic metamaterials}

To demonstrate the effect of the ballistic resonance for designing the optical properties of complex composites, we
consider the planar metamaterial, where  the free electron gas is confined to the conducting layers with the width $d$, surrounded by undoped dielectric ``barriers'' of thickness $D$ (see the schematics in Fig.\ref{fig:2}(a)). In the limit $d,D\ll \lambda$, the multilayer metamaterial behaves as a homogeneous media with uniaxial dielectric permittivity tensor related to permittivities of its components (see {\it Appendix D}) via \cite{Viktor}  
\begin{eqnarray}
\epsilon_n & = & \left[ \frac{\left(\frac{D}{ \epsilon_d} + \frac{d}{\epsilon_n^{\rm eff}}\right)}{d+D}\right]^{-1}, \nonumber \\
\epsilon_\tau & = & \frac{\left( D\epsilon_d + d\epsilon_\tau^{\rm eff}\right)}{d+D},
\end{eqnarray}

Depending on the relative sign of $\epsilon_n$ and $\epsilon_\tau$, the optical response of the metamaterial resembles that of a uniaxial transparent dielectric ($\epsilon_n>0,\epsilon_\tau>0$), an anisotropic reflective metal ($\epsilon_n<0,\epsilon_\tau<0$), or the strongly anisotropic media commonly referred to as a hyperbolic material ($\epsilon_n\cdot \epsilon_\tau<0$) which is able to support the propagation of strongly confined waves and has been demonstrated to be a flexible platform for subwavelength 
focusing \cite{hyper-focusing-Viktor,hyper-focusing-expt}, imaging \cite{hyperlensEN,hyperlensNE,hyperlens-Zhang,hyperlens-3D}, and lifetime and emission manipulation.\cite{ZJISEN,NoginovQED,ShalaevQED} Note that in all previous realizations of hyperbolic metamaterials, the constituent metallic layers have been assumed to operate in the Drude regime. 

The ballistic resonance has profound implications for the electromagnetic response of planar metamaterials. Even if the system is operated at frequencies well above the plasma resonance ($\omega > \omega_p$) such that all constituents of the metamaterial show positive bulk permittivities (${\rm Re}\left[\epsilon_M\right] > 0$, $\epsilon_d > 0$), near the ballistic resonance ${\rm Re}\left[ \epsilon_n^{\rm eff} \right] < 0$, so that the composite behaves as a hyperbolic metamaterial. This offers unprecedented potential in the design and fabrication of plasmonic and hyperbolic metamaterials that do not need any negative-permittivity components.

\begin{figure*}[htbp] 
   \centering
    \includegraphics[width=7. in]{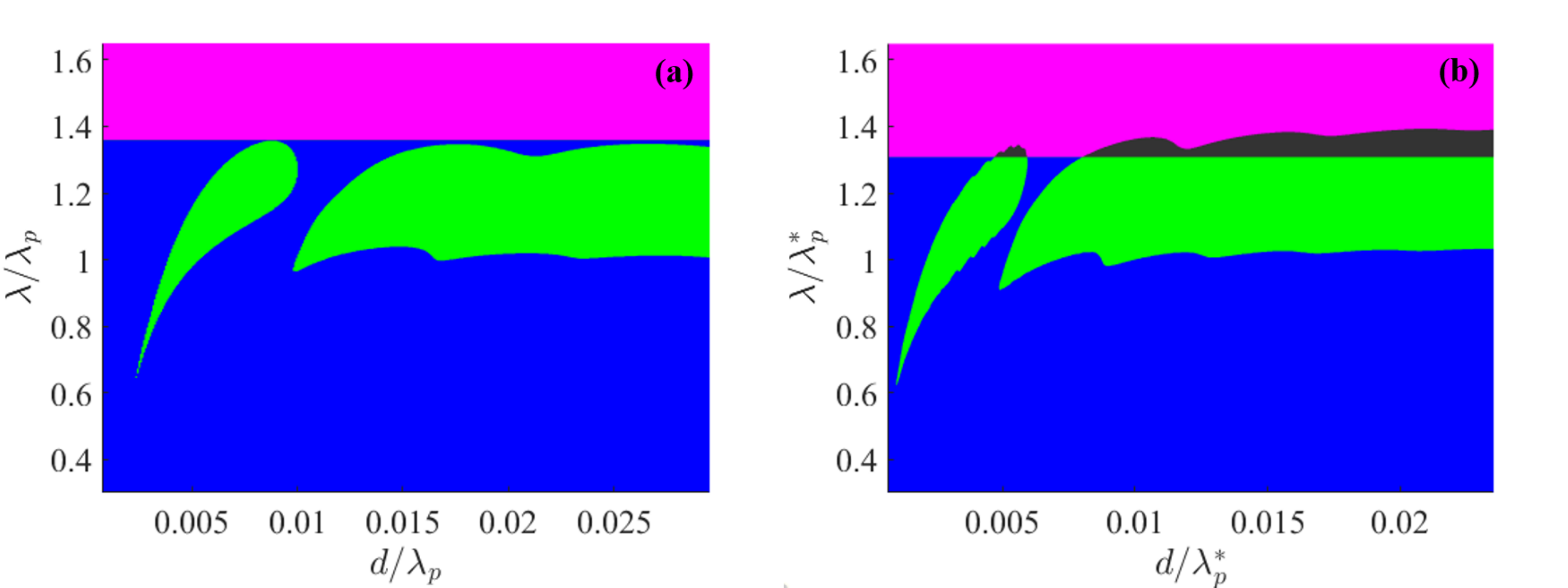}
     \caption{The phase diagram of planar ballistic metamaterials calculated using (a) the semiclassical theory  and (b) the full quantum solutions of the underlying carrier dynamics. As in Fig. \ref{fig:2}, the hyperbolic phase with ${\rm Re}\left[\epsilon_\tau\right] < 0$, ${\rm Re}\left[\epsilon_n\right] > 0$  is represented by magenta color,  the hyperbolic phase  with ${\rm Re}\left[\epsilon_\tau\right] > 0$, ${\rm Re}\left[\epsilon_n\right] < 0$ is shown in green, the elliptic (dielectric) phase -- with blue, and the metallic phase -- with gray color. Note that the semiclassical approach adequately describes the optical response of the composite even in the limit of ultra-thin layers. The slight quantitative disagreement between the predictions of the two techniques is attributed to the fact that the semiclassical formalism neglects the band non-parabolicity effects that are accounted for in the numerical solutions.  }
     \label{fig:qm}
\end{figure*}

Fig. \ref{fig:2} shows the phase diagrams of the planar metamaterial with $d=D$ using the conventional Drude approach
and the ballistic model discussed in this work. Here we see the contrast between the conventional effective medium approach that ignores the inherent mobility of the free charge carriers (panel (b) and the actual behavior (panel (c)). 

In the framework of the effective-medium theory with the local (Drude) permittivities,\cite{ref:nmat,Viktor}   the phase diagram only shows three different ``phases''  of the electromagnetic response of the metamaterial: the dielectric phase with ${\rm Re}\left[\epsilon\right] > 0$  (blue color in Fig. \ref{fig:2}), the type-I hyperbolic  phase \cite{ref:Zubin-types}  with ${\rm Re}\left[\epsilon_n\right] < 0$, ${\rm Re}\left[\epsilon_\tau\right] > 0$ (green color in Fig. \ref{fig:2}), and the type-II hyperbolic phase \cite{ref:Zubin-types}  with ${\rm Re}\left[\epsilon_n\right] > 0$, ${\rm Re}\left[\epsilon_\tau\right] < 0$  (magenta color in Fig. \ref{fig:2}). Note that in the local theory, the 
boundaries between different ``phases"  do not depend on the layer thickness $d$ (as long as the system is within the metamaterial limit $d \ll \lambda$) -- see Fig. \ref{fig:2}(b). 

However, the actual phase diagram (see Fig. \ref{fig:2}(c)) is substantially different from this comparatively simple behavior. For a relatively large value of the layer thickness $d$ (so that $d \gg v_F/\omega$), the system is in the diffusive regime, and its evolution with a change in the frequency is similar to the predictions of the local theory -- see Fig. \ref{fig:2}(b). However, even in this diffusive regime there are visible differences in the phase diagrams. Now, two topologically different hyperbolic phases are no longer connected, but instead separated from each other by a narrow region of the dielectric response -- see Fig. \ref{fig:2}(c).  Furthermore, the frequencies of the phase transitions that define the phase boundaries of the type-I hyperbolic phase (represented by the green color in the phase diagrams)  now nontrivially depend on the layer thickness. This is a direct consequence of the ballistic resonance, whose signatures are already apparent even in the diffusive regime.  Close to the resonance, the free carrier response is enhanced, and the resulting hyperbolic behavior is extended to higher frequencies. However, due to the general properties of periodic resonant response, the resonance frequencies of different orders are separated by ``anti-resonances'' of the reduced response, similar to the constructive and distructive interference in wave phenomena. Close to these ``anti-resonances'', the free-carrier response is suppressed, and the frequency bandwidth of the type-I hyperbolic phase is reduced below the range that is expected from the local theory -- see Fig. \ref{fig:2}(c).

At smaller layer thickness (corresponding to lower orders $n$ in the ballistic resonance condition $d = (2 n + 1) \pi v_F/\omega$) the ballistic effects  become progressively stronger, with increasingly broader frequency range of the hyperbolic response on-resonance and reduced bandwidth at anti-resonance, leading to a complete separation of the hyperbolic phase for $n=0$ into an ``island'' in the phase diagram of Fig. \ref{fig:2}(b), and the type-II hyperbolic phase (magenta color in Fig. \ref{fig:2}). Defined by the zero-order ballistic resonance, this island of type-I hyperbolic response follows the line of $\omega \sim \pi v_F/d$ at progressively higher frequencies, well above the plasma frequency $\omega_p$ -- see Fig. \ref{fig:2}(c). Indeed, the hyperbolic phase now extends to frequencies that substantially exceed the plasma frequency of the conducting layers. 

For the semiclassical description to apply in the ballistic regime,  the layer thickness should exceed the electron de Broglie wavelength at the Fermi energy $\lambdabar_F$, while at the same time be on the order of or smaller than the electron mean free path $\ell$ and the characteristic electron displacement per single electromagnetic period $v_F/\omega$, which implies that
\begin{eqnarray}
v_F / \omega_p \gg \lambdabar_F, \label{eq:sc1}
\end{eqnarray}
where $v_F$  is the Fermi velocity. The inequality (\ref{eq:sc1}) can be expressed in terms of the carrier density $n_e$ as
\begin{eqnarray}
n_e & \gg & n_{sc} \equiv \left( \frac{4}{3^\frac{4}{3} \pi^\frac{12}{5} } \frac{m_* e^2}{\hbar^2 \epsilon_\infty } \right)^3, \label{eq:scn}
\end{eqnarray}
Depending on the background dielectric permittivity (that excludes the contribution of free carriers) $\epsilon_\infty$ and the free carrier effective mass, Eqn. (\ref{eq:scn}) leads to $n_{sc} \sim10^{14} \ {\rm cm}^{-3}$ for a semiconductor from the III-V family such as a gallium arsenide \cite{ref:nmat} ($\epsilon_\infty \simeq 12$, $m_* \simeq  0.06 m_0$), to  $n_{sc} \sim 10^{18} \ {\rm cm}^{-3}$ in conducting oxides such as ITO \cite{ITO}  ($\epsilon_\infty \simeq 4.5 $, $m_* \simeq  0.5 m_0$), and to  $n_{sc} \sim 10^{19} \ {\rm cm}^{-3}$ for plasmonic metals such as gold or silver \cite{AlexModels} ($\epsilon_\infty \simeq 4$, $m_* \simeq   m_0$). For the semiconductor - based metamaterials \cite{ref:nmat} studied in our experiments, the characteristic doping level is at the level of $10^{18} \ldots 10^{19} \ {\rm cm}^{-3}$, and the semiclassical criterion of (\ref{eq:scn}) is therefore well satisfied. In plasmonic metals,  the actual free electron densities  at the level of $n_e \sim 10^{22} \ {\rm cm}^{-3}$ also substantially exceed the corresponding critical value of $n_{sc}$, $n_e \gg n_{sc}$. On the other hand, for conducting oxides the free carrier density is on order of $10^{19} \ {\rm cm}^{-3}$, which is relatively close to the corresponding value of $n_{sc}\sim 10^{18} \ {\rm cm}^{-3}$. As a result, the semiclassical regime, while relatively limited in the metamaterials based on transparent conducting oxides, dominates the response of semiconductor and high-quality single-crystal metal-dielectric metamaterials.

{

However, in absolute terms the ballistic resonance imposes substantially more stringent requirements on the fabrication quality of metal-dielectric composites than on that of the semiconductor metamaterials. Ballistic resonance is rapidly suppressed by the diffuse surface scattering of the free electrons, when the characteristic interface roughness exceeds the electron de Broglie wavelength $\lambdabar_F$. In a plasmonic metal, the latter is on the order of the inter-atomic distance -- so for the ballistic resonance in a metal-dielectric composite one needs atomic level fabrication accuracy. In contrast to this behavior, the de Broglie wavelength in a semiconductor ``designer metal'' is on the order of $10$ - $100$ nm, so interface roughness well under a nanometer typical for semiconductor metamaterials (e.g.  $0.17$ nm in high-quality GaInAs/AlInAs quantum-cascade structures \cite{Tsujino}), will have little effect on the  ballistic resonance.
}

\section*{Topological transitions in Ballistic Metamaterials}

As seen in Fig.\ref{fig:2}, for fixed values of the layer thicknesses, our planar metamaterial exhibits a series of topological transitions between elliptical, hyperbolic, or metallic phases as a function of operating frequency. Each of these transitions has a clear signature in the angle-resolved absorption spectra of the composite. Indeed, the high-frequency transition from elliptic into type-I hyperbolic phase is associated with an absorption resonance that monotonically increases as a function of incident angle (see {\it Appendix A}). At the same time, the transition into the type-II hyperbolic phase is associated with an absorption resonance that monotonically decays as a function of incident angle. 

Importantly, only $p$-polarized light is affected by these topological transitions, since the  propagation of $s$-polarized light only depends on $\epsilon_\tau$.  Therefore, a direct observation of the angle-enhanced absorption peak that is only present in the $p$-polarized response of a planar metamaterial, at a frequency that substantially exceeds the value of $\omega_p$ (which can be measured independently), offers a clear and unambiguous signature of the hyperbolic band induced by the ballistic resonance. 

While the origin of the ballistic resonance lies in the semiclassical dynamics of the system, the observed behavior is consistent with a full quantum description of the metamaterial. In a sense, the ballistic response describes a transition of the behavior of relatively thick ``quantum'' wells with large number of energy levels towards their infinitely thick counterparts, where electrons are best described as free carriers. Fig.\ref{fig:qm} illustrates the comparison between the topological phase diagram of the same metamaterial, as predicted by semiclassical theory and by full quantum mechanical calculations of the optical response of the metamaterial.  Note that the semiclassical theory provides reasonable agreement with the full quantum solutions even in the limit of ultra-thin layers.

 \begin{figure*}[htbp] 
   \centering
    \includegraphics[width=7. in]{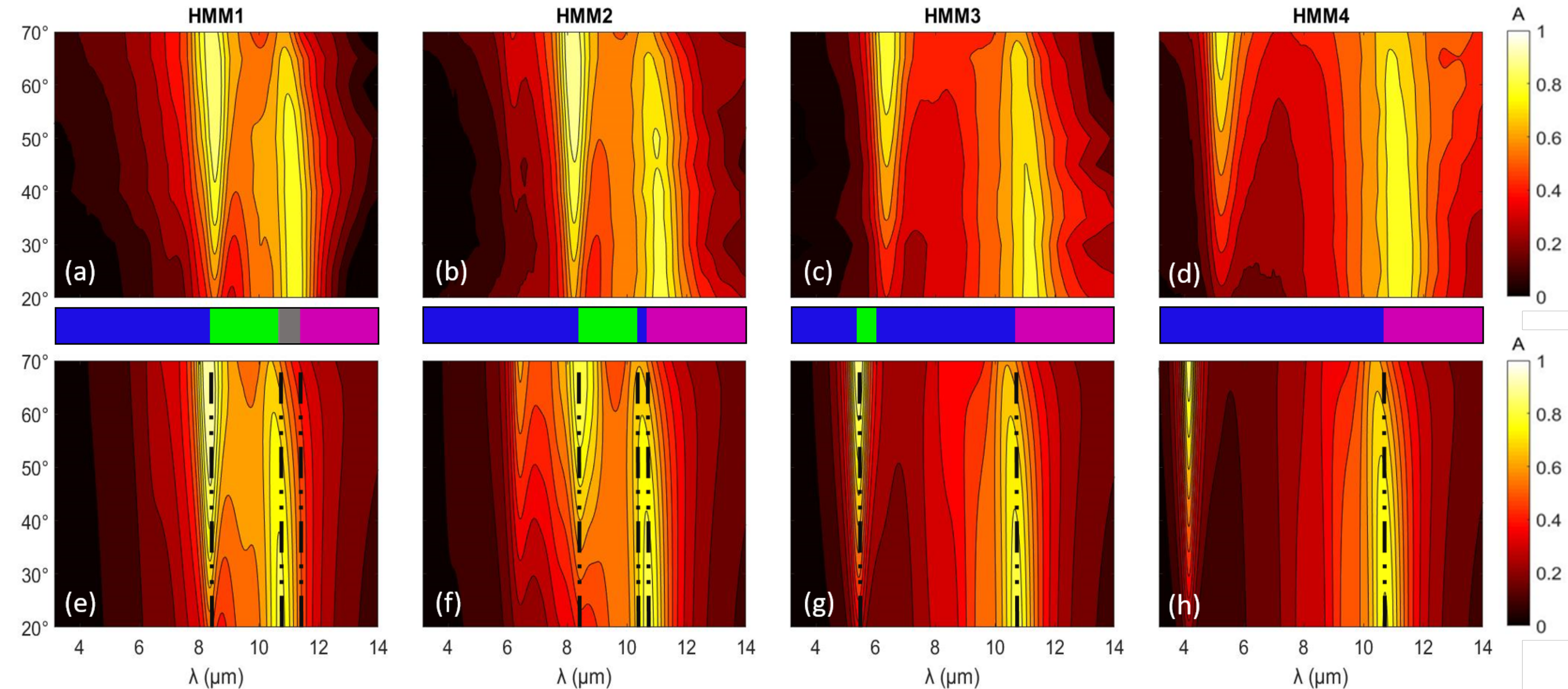}
     \caption{ The absorption spectra of TM ($p$) polarized light incident on $4 \ \mu{\rm m}$-thick semiconductor metamaterial stacks grown  on indium phosphide substrates as a function of incidence angle, for the single layer width $d$ of HMM1: $d = 80$ nm (a,e), HMM2: $d = 33$ nm (b,f), HMM3: $d = 9.5$ nm (c,g), and HMM4: $d = 5.5$ nm (d,h). Panels (a...d) show experimental data, while the panels (e...h) corresponds to the results of the calculation that include the effect of the finite barrier height and band nonparabolicity. Colored insets illustrate optical topology; colors correspond to those used in Fig.\ref{fig:2}. 
     }
     \label{fig:absorption}
\end{figure*}

\section*{Experimental Results}

 For the experimental demonstration of the ballistic metamaterials and the associated resonance, III-V semiconductors  
offer several important advantages. First, as we will show in the subsequent analysis, this material platform \cite{ref:nmat} offers a broad range of the system parameters (such as e.g. the doping density, or the material composition 
in the AlGaInAs alloys) that supports the semiclassical regime that lies in the heart of the ballistic resonance.  In particular, epitaxially-grown III-V alloys allow for doping concentrations high enough for the semiconductor to behave as a plasmonic material at infrared wavelengths.\cite{Law2012} Second, III-V alloys offer high-quality, low material loss composites with near-atomic quality interfaces.\cite{ref:nmat} The high quality of the interface is essential for ballistic metamaterials, as diffuse surface scattering will quickly suppress the ballistic resonance. Finally, with this  choice the resulting metamaterials can share the same material platform with existing optoelectronic devices, which will greatly facilitate the introduction of such metamaterial elements into practical nanophotonic devices.

For the experimental demonstration of the ballistic metamaterials, we therefore use the ``designer metal'' semiconductor 
platform \cite{ref:nmat} where the free electron gas results from the controllable doping of InGaAs layers, surrounded by undoped InAlAs dielectric ``barriers'' (see the schematics in Fig.\ref{fig:2}(a)).  Our semiconductor metamaterial samples were grown lattice-matched to semi-insulating (100) InP substrates and consist of $N$ periods of alternating $n^{++}$ InGaAs and undoped InAlAs layers of equal thickness ($d = D$).  We have grown and characterized four metamaterial samples  with decreasing period ($\Lambda = 2 d$):  HMM1 ($\Lambda = 160$ nm), HMM2 ($\Lambda = 66$ nm), HMM3 ($\Lambda = 19$ nm), and HMM4 ($\Lambda = 11$ nm).  All four samples were grown to have total HMM thickness of approximately 4 $\mu$m.  The details of the sample fabrication are described in {\it Appendix A}.

The doping concentration in our samples was determined by first growing bulk ($500$ nm) $n^{++}$ - doped InGaAs layers on InP substrates.  These bulk, highly doped InGaAs layers were characterized by IR reflection spectroscopy and the resulting spectra fitted using a transfer matrix method approach with the $n^{++}$ InGaAs modelled as a Drude metal, with plasma wavelength ($\lambda_p$) and scattering rate ($\gamma \equiv 1/\tau$) as fitting parameters.  This approach of independent determination of the material parameters  from experiments on the bulk samples allowed us to perform a direct comparison of our theoretical results to the measurements of transmission and reflection from the semiconductor metamaterials HMM1, HHM2, HHM3 and HHM4 with {\it no additional fitting parameters}.

The as-grown metamaterial samples HMM1, HHM2, HHM3 and HHM4 were characterized by angle- and polarization- dependent reflection and transmission spectroscopy using a Fourier transform infrared (FTIR) spectrometer with a custom built external reflection/transmission set-up. The details of the sample characterization and the transmission/reflection measurements are described in {\it Appendix A}.

In Figs. \ref{fig:absorption} and  \ref{fig:7}  we present the comparison of the experimental data for the absorption (Fig. \ref{fig:absorption}) and transmission (Fig.  \ref{fig:7}) with the corresponding theoretical results.  Here, the theoretical calculation includes both the finite value of the conduction band offset between the InGaAs and AlInAs layers and
the band nonparabolicity that becomes substantial at high doping densities. Note the quantitative agreement between the 
theory and the experimental data, with no fitting parameters.\cite{Note1}

The ``smoking gun'' of the ballistic metamaterial behavior is the (single) transmission dip and the corresponding absorption peak at the wavelength that is well below the value of plasma wavelength of the material(s) forming the composite, as seen in Fig.\ref{fig:7} and in Fig.\ref{fig:absorption}. As expected, our transmission data show well-defined transmission dips at the wavelengths of 8.53 $\mu$m (HMM1), 8.25 $\mu$m (HMM2), 
6.35 $\mu$m (HMM3), and 5.25 $\mu$m (HMM4), which correspond to factors 0.996 (HMM1), 1.03 (HMM2), 1.34 (HMM3) and 1.62 (HMM4) smaller than the plasma wavelength (8.5 $\mu$m) of the conducting layers. 

\begin{figure}[htbp] 
   \centering
    \includegraphics[width=3.5 in]{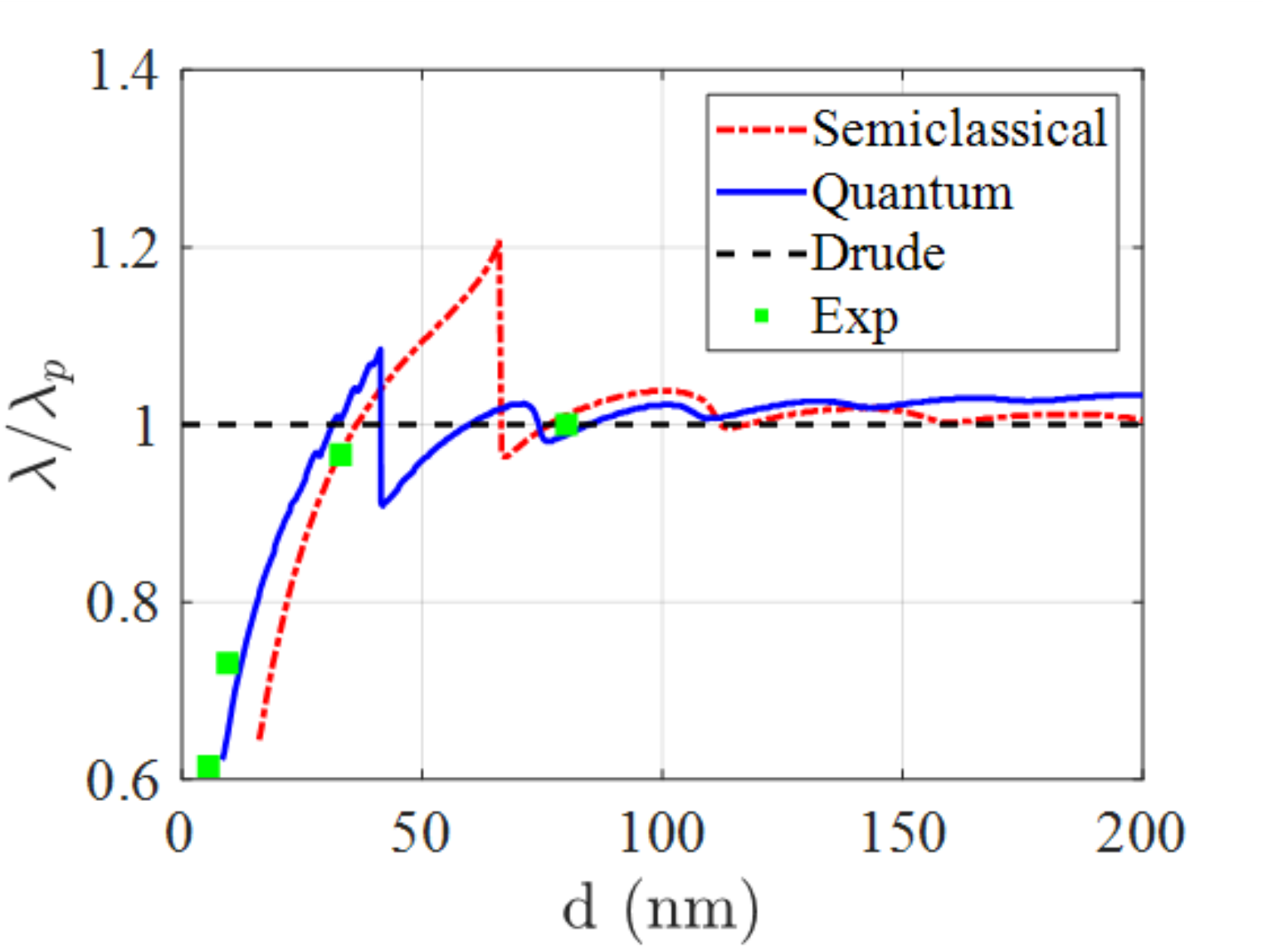}
     \caption{ Dependence of wavelength of the topological transition in ballistic metamaterials on the layer thickness, as predicted by the full quantum mechanical numerical calculations (solid blue line), by the analytical semiclassical theory (dash-dotted red line) and as observed in experiments (symbols). The dashed gray horizontal line corresponds to the conventional (Drude model - based) theory that predicts no variation with $d$. 
     }
     \label{fig:sg}
\end{figure}

Fig.\ref{fig:sg} summarizes these trends by overlapping the experimental data with predictions of both full-quantum and semiclassical theories. The small deviation between the results of the semiclassical and quantum theories shown in Fig. \ref{fig:sg},  is due to band non-parabolicity that is taken into account in our quantum mechanical model but that is neglected by the semiclassical description. Most importantly, our experimental data are in quantitative agreement with both theoretical calculations - without the use of any fitting parameters. 
The increasing blueshift of the measured absorption features align with the signature of the topological transition predicted in our theory and provide conclusive evidence for the ballistic regime in semiconductor metamaterials.

\begin{figure*}[htbp] 
   \centering
    \includegraphics[width=7. in]{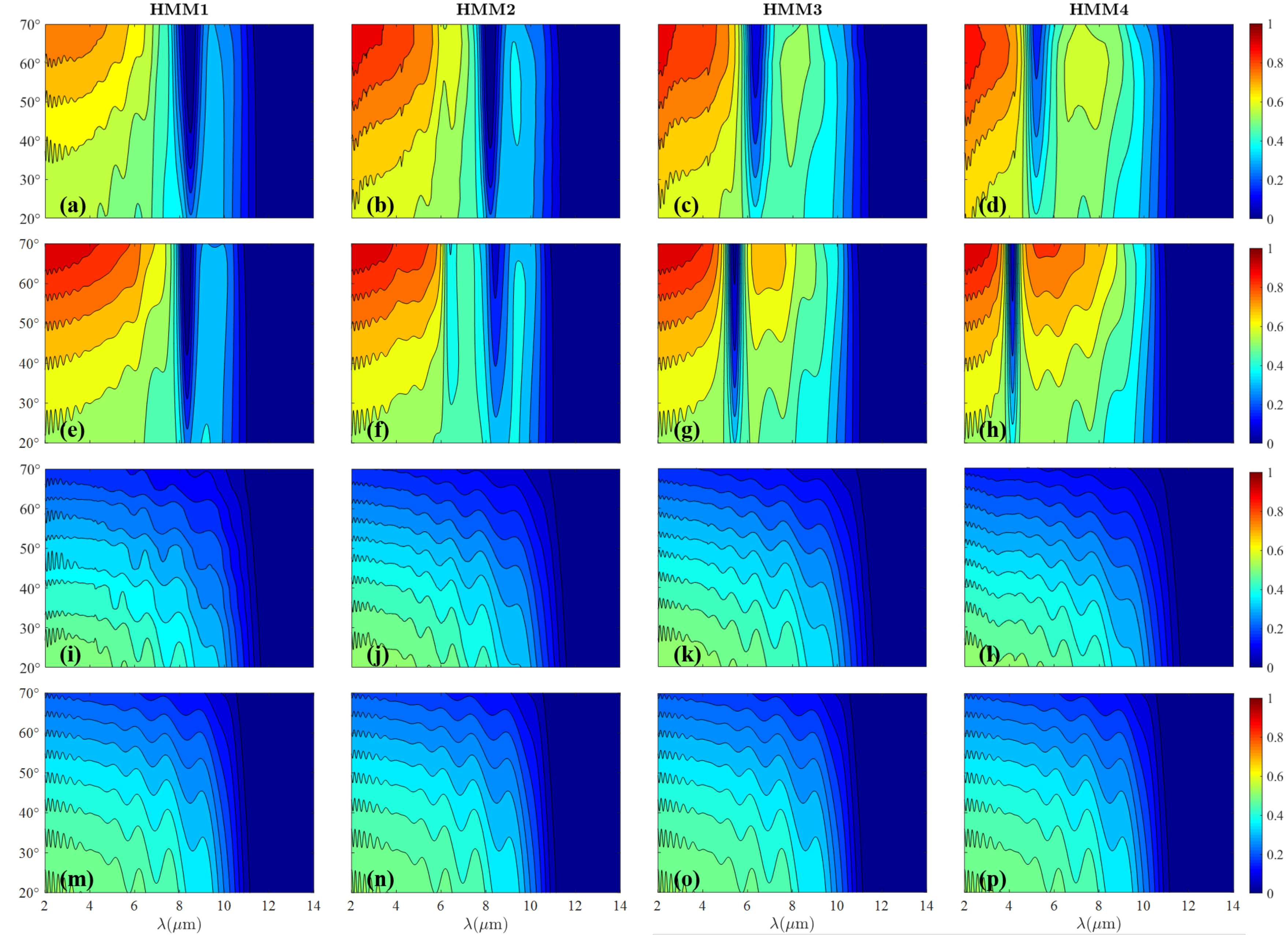}
     \caption{The magnitude of the transmission coefficient in TM ($p$) polarization (a...h) and in TE ($s$) polarization (i...p) (in the false-color representation) for light incident on $4 \ \mu{\rm m}$-thick semiconductor metamaterial stack, fabricated on indium phosphide substrate, vs. incidence angle and the wavelength, for the single layer width $d$ of HMM1: $d = 80$ nm(a,e,i,m), HMM2: $d = 33$ nm (b,f,j,n), HMM3: $d = 9.5$ nm (c,g,k,o), and HMM4: $d = 5.5$ nm (d,h,l,p). Panels (a...d, i...l) show experimental data, while panels (e...h, m...p) correspond to the results of the calculation that include the effect of the finite barrier height and band nonparabolicity.
     }
     \label{fig:7}
\end{figure*}

\section*{The Discussion and Conclusions}

In this work we introduced a new approach to designing the optical response of conducting nanostructures that takes advantage of the inherent mobility of the free electrons in conducting elements of the metamaterial unit cell. We have predicted the phenomenon of ballistic resonance, that leads to the negative dielectric permittivity well above the bulk plasma frequency of the bulk material. Our experimental data offer a direct observation of this new physical phenomenon in multilayered metamaterial, when a composite formed entirely from ``dielectric'' (${\rm Re}\left[\epsilon\right] >0$) materials, shows hyperbolic behavior. 

The ballistic approach to optical metamaterials introduced in our work can therefore dramatically extend the application range of non-metallic material platforms, and open the field of nanophotonics  and nanoplasmonics to materials that until now were considered unsuitable for such applications. In particular, the hyperbolic response of the planar semiconductor 
metamaterials above the plasma frequency of its highly-doped component, demonstrated in our experiments, shows that the use of these composites, originally introduced as ``designer metals''  for mid- and far-infrared frequencies,\cite{ref:nmat} can now be extended closer to their interband absorption edge in the near-IR range. Our analysis suggests that similar phenomena are expected in noble metal platforms, potentially pushing these platforms into UV frequencies. 

Finally, the ballistic resonance presented in this work is expected to play important role in the design of nanostructures with different geometries, including nanowire and nanoparticles, as well as on nanowire- and nanoparticle-composites. 

\section*{Acknowledgements}

This work was partially supported by the National Science Foundation grants DMR-1629276, DMR-1629330, and DMR-1629570, and by Gordon and Betty Moore Foundation.

\begin{appendix}
\section{Sample Fabrication and Characterization}

The hyperbolic metamaterial samples investigated in this work were grown by molecular beam epitaxy (MBE) in a Varian Gen II system with a valved arsenic cracker and effusion sources for gallium, indium and the silicon dopant.  
All samples were grown lattice-matched to semi-insulating (100) InP substrates and consist of multiple periods of alternating $n^{++}$ InGaAs and InAlAs layers of equal thickness $d = 80$ nm (sample HMM1),  $d = 33$ nm (sample HMM2),  $d = 9.5$ nm (sample HMM3), and $d = 5.5$ nm (sample HMM4). In all metamaterial samples, the total thickness of the hyperbolic multilayer was  $\simeq 4 \mu$m.  

The doping concentration was independently determined by first growing bulk (500 nm) $n^{++}$ - doped InGaAs layers on InP substrates.  The bulk, highly doped InGaAs layers were characterized by IR reflection spectroscopy and the resulting spectra fitted using a transfer matrix method approach with the n++ InGaAs modeled as a Drude metal, { which}  lead to the effective
plasma wavelength of $\lambda_p = 8.5 \  \mu$m and the scattering time of  of $\tau \equiv 1/ \gamma = 0.1$ ps. 
In addition, Hall measurements were used to determine the doping concentration of the of the bulk InGaAs ($n_e= 1.78 \cdot 10^{19}$ cm$^{-3}$), and it is this carrier concentration which was used for the numerical simulations modeling the optical behavior of the HMMs studied in this work.

 Samples were characterized by angle- and polarization-dependent Fourier transform infrared (FTIR) reflection and transmission spectroscopy).  Light from the internal FTIR source was collimated using two sequential apertures with diameters of approximately $1$ mm.  A wire grid polarizer between the apertures was used to control the polarization of the incident light.  The sample was mounted on a rotational stage with axis of rotation through the sample.  For transmission experiments, transmitted light was collected by a reflective parabolic lens and focused onto a liquid nitrogen cooled HgCdTe (MCT) detector.  The sample was rotated and spectra were  collected for each polarization (TE and TM).  All transmission spectra were normalized to the spectra collected with no sample in the beam path.  For reflection measurements, the collecting lens and detector  { were}  rotated about the samples' axis of rotation and positioned to collect the specular reflection from the sample.  All reflection spectra  { are}  normalized to the reflection from a gold mirror, which in the mid-IR {has} near unity reflectivity for the angles and polarizations investigated.

\section{The effective medium theory}

In the framework of the effective medium theory, the normal ($\epsilon^{\rm eff}_n$) and the tangential ($\epsilon^{\rm eff}_\tau$) components of the effective permittivity tensor of the conducting layer can be defined as 
\begin{eqnarray}
 \epsilon_{n,\tau}^{\rm eff} & = &  \epsilon_\infty  +  4 \pi  \frac{\langle P_{n, \tau}\rangle }{\langle E_{n, \tau} \rangle }, 
 \label{eq:eps_eff}
\end{eqnarray}
where ${\bf P}$ is the free electrons' polarization induced in the conducting layer in response to the electric field ${\bf E}$, and the brackets $\langle \ldots \rangle$ represent the average over the thickness of the conducting layer $d$:
\begin{eqnarray}
\langle \ldots \rangle & \equiv & \frac{1}{d} \int_0^d  \ dz \ldots
\label{eq:average}
\end{eqnarray}
Alternatively, the normal component of the effective permittivity can be defined in terms of the ratio of the average field in the layer 
$\langle E_n \rangle$ to the field at its boundary $E_n(+0)$:
\begin{eqnarray}
\epsilon^{\rm eff}_n & = & \epsilon_\infty + \frac{\langle E_n\rangle}{E_n(+0)}.
\label{eq:en_eff}
\end{eqnarray}
When the field in the conducting layer ${\bf E}$ and the free electron polarization ${\bf P}$ satisfy Gauss's law
\begin{eqnarray}
\epsilon_\infty \ {\rm div}{\bf E} & = & 4 \pi {\bf P},
\end{eqnarray}
the two definitions of the effective permittivity (\ref{eq:eps_eff}) and (\ref{eq:en_eff}) are equivalent to each other.

The problem of finding the electromagnetic response of a metamaterial in the ballistic regime therefore reduces to the task of finding the polarization of a single conducting layer in the external field.

\section{Semiclassical approach.} 

In the semiclassical approximation, the free carrier response can be treated within the semiclassical framework,
via the Boltzmann kinetic equation  \cite{Ziman} :
\begin{eqnarray}
\frac{\partial f_{\bf p}}{\partial t}  + {\bf v}_{\bf p} \cdot \nabla f_{\bf p} +
 e {\bf E}\cdot {\bf v}_{\bf p}  \frac{\partial f_0}{\partial \varepsilon_{\bf p}} & = & - \frac{f_{\bf p}  - f_0}{\tau},
\label{eq:Boltzmann}
\end{eqnarray}
where $f_{\bf p}\left({\bf r},t\right)$ is the charge carriers' distribution function with its equilibrium (Fermi-Dirac) limit $f_0$, $\varepsilon_{\bf p}$ is the electron energy for the (Bloch) momentum ${\bf p}$,  ${\bf v_p} \equiv \partial\varepsilon_{\bf p}/\partial{\bf p}$ is the corresponding electron group velocity, and $\tau$ is   the effective relaxation time defined by the bulk scattering (due to e.g. phonons, impurities, etc.) The local equilibrium distribution function $f_0$   is defined by the actual time-dependent local density  rather than its time-averaged value,\cite{KraglerThomas1980} if the scattering process does not locally create or annihilate charge carriers. However, when the electromagnetic field frequency $\omega \gg 1/\tau$, the local correction to the equilibrium distribution function can be neglected.\cite{KliewerFuchs1968,MelykHarrison1970}

For a high-quality interface along one of the symmetry planes of the crystal, the surface leads to specular reflection of the charge carriers,\cite{Ziman} which can be accounted for by the boundary condition on the distribution function,
\cite{Ziman,Fuchs1938,ReuterSondheimer1948,Sondheimer1950,Soffer1967}
\begin{eqnarray}
f_{{\bf p}^-}\left({\bf r}_s\right) & = &f_{{\bf p}^+}\left({\bf r}_s\right),
\label{eq:ss_specular}
\end{eqnarray}
where ${\bf p}^+$ and ${\bf p}^-$ are connected by the specular reflection condition, and  ${\bf r}_s$ corresponds to any point at the interface.

Self-consistent solution of the kinetic equation (\ref{eq:Boltzmann}) with the boundary condition (\ref{eq:ss_specular}) together with Maxwell equations, where the charge density $\rho\left({\bf r},t\right)$ and the current density ${\bf j}\left({\bf r}, t\right)$ of the free electrons are defined by the distribution function $f_{\bf p}\left({\bf r},t\right)$,
\begin{eqnarray}
\rho\left({\bf r}, t\right)  & = & 2 e \int \frac{d{\bf p}}{\left(2 \pi \hbar\right)^3} \cdot \left(f_{\bf p}\left({\bf r}, t\right) - f_0\left(\varepsilon_{\bf p}\right) \right), \\
\label{eq:rho}
{\bf  j}\left({\bf r}, t\right)  & = & 2 e \int \frac{d{\bf p}}{\left(2 \pi \hbar\right)^3}\  {\bf v}_{\bf p}   \  \left(f_{\bf p}\left({\bf r},t \right) - f_0\left(\varepsilon_{\bf p}\right) \right)
\label{eq:j}
\end{eqnarray}
yields the actual spatial distribution of the electric field and the free electron polarization in the conducting layer, which then define the effective permittivity tensor via Eqns. (\ref{eq:eps_eff}) or  (\ref{eq:en_eff}). 

\section{Quantum theory: von Neumann - Maxwell problem.} 

In the full quantum-mechanical description of the electromagnetic response of the conducting layer, the free carrier contributions to the electronic polarization can be expressed in terms of the electron density matrix $\hat{\rho}$ and the corresponding charge density and current density operators,
\begin{eqnarray}
\rho\left({\bf r}, t\right) & = &  e \ {\rm Tr}\left[ \delta\left(\hat{\bf r} - {\bf r}\right) \hat\rho\right]  \\ \label{eq:rhomtr}
{\bf j}\left({\bf r}, t\right) & = & \ {\rm Tr}\left[  \hat{\bf j}\left({\bf r}\right)  \hat\rho\right].   \label{eq:jmtr} 
\end{eqnarray}
Here, the (time-dependent) density materix is defined by the standard von Newmann - Liouvillie equation
\begin{eqnarray}
i\hbar \frac{\partial\hat{\rho}}{\partial t} = \left[\hat{H}, \hat{\rho}\right] - \frac{\hat{\rho} - \hat{\rho}_0}{\tau},
\end{eqnarray}
where $\hat{\rho}_0$ is the free electron density matrix at the equilibrium, and the Hamiltonian $\hat{H}$ includes the contributions 
of the electromagnetic field (via the standard scalar and vector potentials formalism). 

Self-consistent solution of the von Neumann-Liouville equation and the Maxwell's equations with the charge and current densities expressed in terms of the free electron density matrix, yields the distribution of the electromagnetic field and the free carrier polarization in the conducting layer, which then via Eqns. (\ref{eq:eps_eff}) or  (\ref{eq:en_eff}) define the effective permittivity.  

In the limit of large bandgap in the dielectric that surrounds the conducting layers (so that the exponential ``tail'' of the free electron wavefunction entering the dielectric can be neglected), we obtain
\begin{eqnarray}
\epsilon^{\rm eff}_n & = & \epsilon_\infty \left\{ 1 + \frac{ \omega_p^2}{ \left(\omega + i / 2\tau\right)^2} \right.  \nonumber \\
& \times &\left.  G\left(N_F, 
\frac{2 m_* d^2}{\pi^2 \hbar} \left(\omega + i/2\tau  \right)\right) \right\},
\label{eq:enq} 
\end{eqnarray} 
where
\begin{eqnarray}
G\left(N_F, \Omega\right) &  = & -1 + \frac{4 \left(N_F^2 - 1\right)}{n_D d^3 \Omega^2} \sum_{m = 1, \pm}^{\left[ N_F \right]}   m^2 
\cdot 
\sqrt{m^2 \pm \Omega}
 \nonumber \\
& \times & \tan\left[ \frac{\pi}{2} \sqrt{m^2 \pm \Omega} - \frac{\pi}{2} m  \right], \ \ \ \ \ \ \ \ \ 
\label{eq:G}
\end{eqnarray}
$n_D$ is the doping density, $\left[ N_F \right]$ represents the integer part of $N_F$, which is defined by the doping
density by the equation
\begin{eqnarray}
n_D & = & \frac{\pi  \left[ N_F \right]}{2d^3} \left(N_F^2  - \frac{\left[ N_F \right] ^2}{3} - \frac{\left[ N_F \right]}{2} - \frac{1 }{6} \right). 
\label{eq:nD} 
\end{eqnarray}

However, for the quantitative description of the experimental data the finite barrier height in the dielectric layers has to be accounted for. Furthermore, at the relatively high doping of our samples, the conduction band non-parabolicity becomes significant and also needs to be included in the theoretical description.

 \begin{figure*}[htbp] 
   \centering
    \includegraphics[width=7. in]{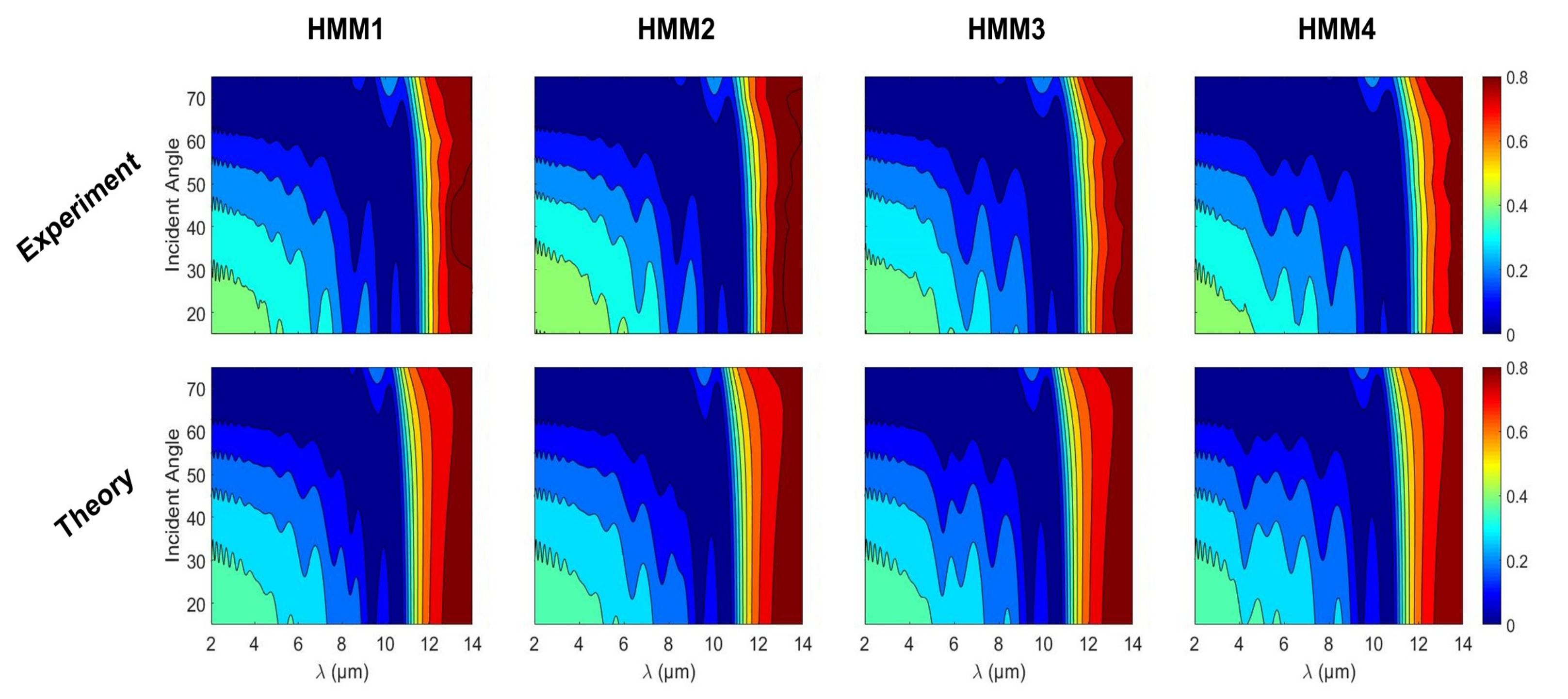}
     \caption{The magnitude of the reflection coefficient in TM ($p$) polarization (in the false-color representation) for light incident on $4 \ \mu{\rm m}$-thick semiconductor metamaterial stack, fabricated on indium phosphide substrate, vs. incidence angle and the wavelength, for the single layer width $d$ of (from left to right)  $80$ nm, $33$ nm, $9.5$ nm and $5.5$ nm. The top row shows experimental data, while the bottom row corresponds to the results of the calculation that include the effect of the finite barrier height and band nonparabolicity.
     }
     \label{fig:8}
\end{figure*}

\section{Band non-parabolicity and finite barriers} 

The effects of the band non-parabolicity and finite conduction band discontinuity at the interface between highly doped and undoped layers are both incorporated in our numerical calculations of permittivity of confined plasma. Specifically, we follow a standard 
approach \cite{Bastard} and represent the total energy of the free charges as the sum of the confinement energy $E_z$ and the kinetic energy of their motion perpendicular to the growth direction ($E_\perp$)  
\begin{eqnarray}
E & = & E_z+ E_\perp
\end{eqnarray}
and utilize linear energy-dependent effective mass inside doped InGaAs: 
\begin{eqnarray}
   m_*(E_z) = m_0(\alpha E_z + \beta)
\end{eqnarray}
with $m_0$ being the mass of the electron, and parameters $\alpha = 0.052$ and $\beta = 0.042$. This model is coupled with BenDaniel and Duke boundary conditions\cite{PRL:V152 1966} at the InGaAs/InAlAs interface (we use energy-independent effective mass $m_* = 0.075 m_0$ inside undoped InAlAs and assume  conduction band offset $0.52$ eV between the two materials). The resulting  numerical solutions yield a set of the bound energy levels and wave-functions representing the charge density distribution across the potential well, parameterized by $k_\perp = \sqrt{2 m_* E_\perp} / \hbar$. 

Once the ($k_\perp$ - dependent) energy levels are calculated, the resulting Fermi energy is determined by enforcing the consistency between numerically-calculated total free-charge density $n = \int n_{1D}(k_\perp)k_\perp dk_\perp$ and experimentally-derived doping levels. The same calculation provides the $k_\perp$-dependent distribution of the linear density of the free charges. 

Finally, effective permittivity of the individual layer is given by 
\begin{eqnarray}
    \epsilon^{\rm eff}_n(\omega) & = & \epsilon_{\infty}\left[1+ \frac{ e^2}{m_*\left(E_f\right) \ \epsilon_\infty} \cdot \sum_j   f_j \cdot  \int_{0}^{\infty} dk_\perp \  k_\perp \right. \nonumber \\ & \times & \left. \  \frac{n_{1D}\left(k_\perp\right)  }{\omega_j^2\left(k_\perp\right)-\omega^2-i\omega\gamma_{j}} \right],
\end{eqnarray}
where the oscillator strength \cite{Boyd} is defined as 
\begin{eqnarray}
f_j & = & \frac{2 \ m_*\left(E_f\right)\ \omega_{j}}{\hbar} \ \left|z_{j}\right|^2, 
\end{eqnarray}
$\omega_j(k_\perp)$ represents the frequency of the transition from below to above the Fermi energy at the given value of $k_\perp$, $\gamma_j$ and $z_j$ describe the inelastic loss and the dipole moment associated with this transition, and $e$ represents electron charge.

The superlattice-induced energy level broadening \cite{KrongPenney}, which is not important for the samples considered in this work, can also be incorporated into the model presented here.

\section{Scaling behavior at ballistic resonance.} 

Note the difference in the frequency profile of the ballistic resonance shown in Fig. \ref{fig:1}(b) from the usual Lorentzian shape: here it's the real part of the permittivity which peaks at the resonance, while the imaginary component that accounts for loss, rapidly drops to near-zero in a step-like fashion.

The  origin of this behavior can be uncovered at the low loss limit, $1/\tau \to 0$, when the function $F_z(x)$ can be expressed as
\begin{eqnarray}
{\rm Re}\left[F_z\left(x\right)\right] & = & \frac{3 x^3}{\pi^5 \left(n_x + \frac{1}{2}\right)^5} \cdot \log\left[\sin\left(\delta x\right)\right] 
\nonumber \\
& - & 3 x^3 \cdot \delta F\left(\delta x\right), \\
{\rm Im}\left[F_z\left(x\right)\right] & = & \frac{3}{\pi^4} \  x^3  \sum_{k = n_x+1}^\infty \frac{1}{\left(k-1\right)^5},
\end{eqnarray}
where $n_x$ is the integer part of the ratio of $x$ to $\pi$,  
\begin{eqnarray}
n_x \equiv \left[\frac{x}{\pi}\right],
\end{eqnarray}
$\delta x$ is the deviation from the ballistic resonance,
\begin{eqnarray}
\delta x \equiv x - \pi \left( n_x + \frac{1}{2}\right), 
\end{eqnarray}
and the function $\delta F$ is real and continuous,
\begin{eqnarray}
\delta F\left(\delta x\right) & = & \int_{\left|\delta x\right|}^{\frac{\pi}{2}} dt \ \cot\left(t\right) \ \left( \frac{1}{\pi^5 \left(n_x + 1/2\right)^5}
\right. \nonumber \\
& -  &  \left. \frac{1}{\left(\pi \left(n_x + 1/2\right) + t \right)^5} \right) - 
\nonumber \\
& - &  \Theta\left(- \delta x\right)   \cdot   \int_0^{\ - \delta x} dt \ \cot\left(t\right) \nonumber \\
& \times &   \left(\frac{1}{\left(\pi \left(n_x + 1/2\right) - t \right)^5} - \frac{1}{\left(\pi \left(n_x + 1/2\right) + t \right)^5} \right) \nonumber \\
& - & \sum_{k=1}^\infty \int_0^{\frac{\pi}{2} } dt \ \cot\left(t\right) \  \left(\frac{1}{\left(\pi \left(n_x + k +  1/2\right) - t \right)^5} \right. 
\nonumber \\
& -  & \left. \frac{1}{\left(\pi \left(n_x + k + 1/2\right) + t \right)^5} \right) , 
\end{eqnarray}
where $\Theta\left(t\right)$ is the Heavyside's unit step function.\cite{Morse} Such discontinuity in the free electron absorption is the unambiguous feature of the Landau damping.\cite{LL}

For a finite value of the electron scattering time $\tau$, the step-like discontinuity (see Fig. \ref{fig:1}(b)) in the imaginary parts of the function $F_z$ and, consequently the dielectric permittivity $\epsilon_n^{\rm eff}$, is replaced by a smooth function
that behaves as $\arccos\left( \delta\omega / \sqrt{\delta\omega^2 + 1/\tau^2}\right)$,  where $\delta\omega$ is the frequency deviation from the ballistic resonance.

\end{appendix}



\end{document}